\documentstyle[amssymb,aps,epsfig,prb]{revtex}

\def\br{{\bf r}}

\def\bG{{\bf G}}

\def\bl{{\bf l}}
\def\bG{{\bf G}}
\def\mathrm{}

\newcommand{\scalegraphlabel}[4]{
\begin{figure}[htbp]
\epsfxsize=#2
\centerline{\epsfbox{#1}}
\caption{#3}
\label{#4}
\end{figure}
}

\begin{document}
% \draft
\twocolumn[\hsize\textwidth\columnwidth\hsize\csname @twocolumnfalse\endcsname

\title{Supercell technique for total-energy calculations of finite charged and polar systems}

\author{M. R. Jarvis, I. D. White}

\address{Theory of Condensed Matter Group, Cavendish Laboratory,
Madingley Road, Cambridge CB3~0HE, United Kingdom}

\author{R. W. Godby} 

\address{Department of Physics, University of York, Heslington, York
YO1~5DD, United Kingdom}

\author{M. C. Payne}

\address{Theory of Condensed Matter Group, Cavendish Laboratory,
Madingley Road, Cambridge CB3~0HE, United Kingdom}

\date{\today}
\maketitle

%============================================================
\begin{abstract}

We study the behaviour of total-energy supercell calculations for dipolar molecules and charged clusters. Using a cutoff Coulomb interaction within the framework of a plane-wave basis set formalism, with all other aspects of the method (pseudopotentials, basis set, exchange-correlation functional) unchanged,  we are able to assess directly the interaction effects present in the supercell technique. We find that the supercell method gives structures and energies in almost total agreement with the results of calculations for finite systems, even for molecules with large dipole moments. We also show that the performance of finite-grid calculations can be improved by allowing a degree of aliasing in the Hartree energy, and by using a reciprocal space definition of the cutoff Coulomb interaction.

\end{abstract}

]
\narrowtext

\section{Introduction} 

Density-functional theory (DFT), typically performed within a local approximation for exchange and correlation such as the local density approximation (LDA), has proved to be an extremely powerful and successful method for the calculation of ground-state properties of electronic systems. The Kohn-Sham formalism \cite{KOS65} allows the ground-state electronic density and total energy to be found through minimisation with respect to a set of one-electron orbitals. Expansion of these orbitals in terms of a plane-wave basis set, in conjunction with the use of pseudopotentials\cite{pseudo} to represent nuclei and core electrons, provides a systematic route to convergence for such calculations and also allows the use of Fast Fourier Transforms (FFTs). Plane waves also make the calculation of forces for molecular dynamics straightforward.

While the use of a plane-wave basis set is most natural for infinite periodic systems such as bulk solids, the computational advantages this offers apply equally for calculations of finite systems (molecules) or systems which are not infinite in all three dimensions (polymers or surfaces). However, as the use of a discrete set of plane waves corresponds to employing periodic boundary conditions (PBCs), the calculation will still be essentially for an infinite system, with the supercell, containing the molecule, periodically repeated. These images will see each other through the long-ranged Coulomb interaction, thus introducing differences from the isolated system.

In the case of a charged finite system, the supercell approach gives a divergent Coulomb repulsion energy and so this interaction must be removed in some way, or alternatively a fully finite calculation undertaken. A neutral system with a dipole (or higher order) moment will also have contributions to the total energy from the interaction between supercells: these will fall off with supercell size, but there is only power-law convergence in the energy, and filling cells with extra vacuum may be impractically expensive. Makov and Payne \cite{guy} developed separate corrections to the total energy functional for the cases of charged and dipolar systems, which improved the convergence of energy differences for test systems. 

In order to avoid the enforced periodicity of a supercell approach, many calculations for molecules and clusters are performed on a finite grid. In the case of a plane-wave basis this requires the use of computationally expensive double-sized FFTs to perform the finite convolution for the Hartree potential. Alternatively, localized Gaussian basis sets are extensively employed in quantum chemistry, but these are harder to implement and do not allow for such straightforward systematic convergence procedures as plane waves.

Finite-size effects are also relevant in other types of many-body calculations. Modified Coulomb interaction techniques have been used for {\it GW} many-body perturbation theory calculations \cite{gwclusters} and Quantum Monte Carlo (QMC) calculations,\cite{fraser,andrew} and these authors proposed that a similar approach could be used for DFT studies of finite systems. However, the physical consequences of using a finite cell to approximate an infinite one differ, as the finite-size effects take the form of spurious correlations in QMC or screening from the contents of other cells in {\it GW} work, and thus it is by no means clear that the relative importance of removing the electrostatic interaction of the ground-state densities in DFT total-energy calculations of finite systems will be as large.

In this work we study charged and dipolar systems using a Hamiltonian with a cutoff Coulomb interaction, and compare with the results of a conventional supercell calculation. As we implement our modification for finite systems within the framework of a pseudopotential plane-wave total-energy method (in this case {\small CASTEP} \cite{castep}), we are able to assess the relative performance, and specifically convergence, of these approaches directly, and investigate in which circumstances a fully isolated calculation is required.  Using the test cases of a $\mathrm{Mg}^{+}$ ion and a NaCl bond we demonstrate that our method removes all interaction between supercells. We then study the cases of a ${\mathrm{Na}_{2}}^{+}$ cluster and the polar molecules dimethylsulphoxide and {\it para}-nitroanisole, which provide some idea of the error involved in unmoderated supercell calculations for realistic charged and polar finite systems. We find that there is almost total agreement in the ground-state geometry determined by the supercell method and for the isolated system, even for the smallest supercell sizes which can contain the density, though it has been suggested in the literature \cite{chelikowsky} that supercell calculations under these circumstances require substantial extra vacuum for convergence. All calculations were performed within the LDA (although our conclusions apply for any local approximation to exchange and correlation), and sampled the Brillouin Zone at the $\Gamma$-point only. We use atomic units throughout.

\section{Computational Method}

In this section we describe how to remove any interaction between supercells in a plane-wave basis set calculation. The total energy used in DFT calculations is 

\begin{eqnarray}
{E[n]} &=& T_{s}[n] + \int V_{\mathrm{ion}}(\br) n(\br) d\br +  E_{\mathrm{ion}}( {\{{\bf R}_{i}\}})\nonumber \\
 &+& \frac{1}{2} \int n(\br) n(\br') V(\br,\br') d\br d\br' + E_{\mathrm{xc}}[n].
\end{eqnarray}

The kinetic energy $T_{s}[n]$ and exchange-correlation energy $E_{\mathrm{xc}}[n]$ are in essence local (i.e., do not interact across cells) and therefore do not need to be changed.\cite{kenote} The terms through which the contents of different supercells interact are the Hartree energy, the ionic potential $V_{\mathrm{ion}}$ and the ion-ion Coulomb repulsion energy $E_{\mathrm{ion}}$. In an infinite crystal these three terms give rise at reciprocal lattice vector $\bG=0$ to infinite contributions\cite{IZC79} which can be shown to cancel.

We first consider the electron-ion term. The ionic potential felt by the electrons is given by

\begin{equation}
V_{\mathrm{ion}}(\br)= \sum_{\bl} \sum_{\alpha} V_{\alpha}(\br + \bl)
\end{equation}
where the sums are over all lattice vectors $\bl$ in the infinite crystal and ions $\alpha$ in the unit cell, and $V_{\alpha}$ is the pseudopotential for atom $\alpha$. In order to carry out a calculation with the Coulomb interaction cut off on the supercell, we simply require the sum over the contents of the unit cell only. Whilst in the infinite case the sum over unit cells means that we require the local pseudopotential only at the reciprocal lattice vectors (all other components in the sum cancel), we now need to integrate the local potential in reciprocal space up to the energy cutoff and then FFT to the correct real-space ionic potential for the finite cell. We use a localized real-space representation of the non-local part of the pseudopotential,\cite{king} which is readily modified to prevent any interaction across cells.

The ion-ion energy is now evaluated easily by a direct sum over the ions in one cell (as opposed to the usual Ewald summation \cite{ewald}).

The remaining term through which supercells interact is the Hartree potential, given by

\begin{equation}
V_{\mathrm{H}}(\br) = \int_{\mathrm{all~space}} \frac{n(\br')}{|\br-\br'|} d\br'.
\end{equation}
As this is an infinite convolution it is normally evaluated in reciprocal space where 

\begin{equation} \label{reciphartree}
V_{\mathrm{H}}(\bG) = n(\bG)V(\bG)
\end{equation}
where $V(\bG)$ is the Fourier transform of the Coulomb interaction, $V(\bG)= 4\pi/|\bG|^{2}$. 

To remove electrostatic interaction between electronic charge densities in different supercells we require the modified Hartree potential,

\begin{equation} \label{finhartree}
V_{\mathrm{H}}^{\mathrm{cut}}(\br) = \int_{\mathrm{SC}} \frac{n(\br')}{|\br-\br'|} d\br'
\end{equation}
where the integral is performed only over the cell containing $\br$. This effectively defines a modified Coulomb interaction,

\begin{equation} \label{vcut}
V^{\mathrm{cut}}(\br,\br') = \left\{\matrix{ & \displaystyle{\frac{1}{|\br-\br'|}} && \mathrm{for} \ \br,\br' \  \mathrm{in \ same \ cell} \cr 
 & 0 && \mathrm{otherwise.}}\right.
\end{equation}
This cutoff can be imposed either in real or reciprocal space, and we test the effect of each in the next section. Such an interaction is clearly no longer a function of $\br-\br'$ only though, and hence does not preserve the simple multiplicative form of (\ref{reciphartree}). Onida {\it et al.} employed a spherically cutoff Coulomb interaction, 

\begin{eqnarray}
V(\br,\br') =  \left\{\matrix{& \displaystyle{\frac{1}{|\br-\br'|}} && \mathrm{for} \ |\br-\br'| < R  \cr
 & 0 &&  \mathrm{otherwise}}\right.
\end{eqnarray}
for {\it GW} calculations \cite{gwclusters} on negatively charged clusters to prevent screening by periodic images in other supercells. As noted in their work, though, an interaction cut off only as a function of $|\br-\br'|$ clearly in principle requires additional separation of densities in adjacent supercells to remove all interaction effects.  

An interaction with the form of (\ref{vcut}) can still be used in conjunction with the convolution theorem, by padding the density with an external region of zero density.\cite{gunathesis} If we consider, for simplicity, a cubic supercell of side $L$, then we take a new grid of side 2$L$, defining the density to be zero on the extra points and evaluating the usual real-space Coulomb interaction on the expanded grid. It can then be seen that performing an FFT on the padded density and interaction, multiplying them as in  (\ref{reciphartree}) and reverse-transforming to the padded cell allows us to extract the correct Hartree potential [as defined in  (\ref{finhartree})] in the original real-space cell. For the numerical transform of the Coulomb interaction the singularity at $\br=\br'$ is integrated and averaged over one grid unit.\cite{singnote} A spherical cutoff in the Coulomb interaction at a range of $\surd 3$ $L$ \ (to span the cell) facilitates an analytical transform to reciprocal space but also means that the padded grid must now have side $(1+\surd 3)$ $L$ \ to strictly avoid any spurious contributions to the Hartree integral.

This approach to cutting off supercell interaction in a plane-wave calculation thus leads to a procedure for performing the Hartree integral identical to that employed in some finite-grid calculations, e.g., for molecular dynamics of charged clusters.\cite{guna} Viewing this procedure as a physical truncation of the Coulomb interaction in an infinite calculation, rather than as a mathematical tool to perform a finite integral, makes the potential for savings in computational effort more apparent. In Fig. \ref{padding} we show electron density in a unit cell of side $L$ \ padded by a distance $D$ \ and interacting with all density within a box of side 2$R$. If $D$ \ and $R$ \ are set equal to $L$ \ then we obtain the correct answer for the finite Hartree energy. If $R$ \ is reduced this corresponds to cutting off interaction between electronic density on opposite sides of the same cell. $D$ \ can always be reduced to $R$, but if it is reduced still further then spurious interaction between density on the edges of adjacent supercells is introduced. As the density will be exponentially small at the box edges for a converged calculation, it is conceivable that reducing $R$, or $D$ \ below $R$, will only introduce very small negative and positive errors respectively in the total energy. As this would be a useful procedure in cases where calculation of the Hartree energy dominates computation time, we test the extent to which such savings can be made in the next section.

Through these relatively straightforward alterations to the standard infinite code, a finite calculation can be performed with all electrostatic interaction between supercells removed, and with all other aspects of the method (pseudopotentials, conjugate gradients, geometry optimisation) unchanged.

\section{Results}

\subsection{Test systems}

We first examined the test case of a charged system, taking the ionisation potential of a Mg atom as used by Makov and Payne to assess the convergence of their energy-correction method. The cutoff Coulomb interaction allows a charged system to be treated in exactly the same way as a neutral system. For a usual supercell calculation it is necessary to insert a neutralising background to remove the divergent repulsion energy of the charged lattice of cells. The binding energy between this background and a lattice of point charges is then accounted for by the Madelung correction, whilst Makov and Payne gave an expression for a further correction in terms of the charge density profile. 

For the neutral atom, use of the cutoff interaction gave almost identical (difference of order $10^{-5}$ eV) results to the unmoderated supercell calculation for all box sizes above 8 {\AA}\, as expected: the Hartree, electron-ion and ion-ion energy terms for the finite and repeated supercell systems sum to the same total in this case, as the atom has no non-zero moments.

The calculation for the positive ion with the cutoff interaction can be undertaken simply by reducing the electron number by one as all interaction between supercells has been removed. This gave an energy for the ion which was converged within 0.01 eV at a box size of 8 \AA, as opposed to 10 \AA\ for equivalent convergence of the energy-correction method. We note that there is no power-law behaviour in the convergence of our calculation, and the answer is in principle exactly that of the true finite system once the box is large enough for the electron density to be zero at the edges.

We next consider a system with a dipole moment. In Fig. \ref{NaCl} we show the potential energy of stretching a NaCl molecule by 0.3 \AA\ from its calculated equilibrium length of 2.231 \AA, as a function of cubic box size used for the calculation. The removal of all electrostatic interaction between supercells eliminates the slow power-law convergence, and also represents some improvement over the convergence with respect to supercell size achieved by the energy-correction approach.

We also tested the effect of reducing the interaction range $R$ \ and the padding distance $D$ \ as outlined in the previous section. At a cubic box size of side 12 \AA\ the total energy for the Mg atom is converged to within a few meV. We found that the range parameter could be reduced below 0.6$L$, and the padding distance then further reduced below 0.3$L$, without either change affecting the total energy by more than $10^{-3}$ eV. As this gives a padded grid of side 1.3$L$ \  (as opposed to 2$L$  \ in the exact unaliased case) this can actually give an improvement of almost an order of magnitude in the time taken to compute the Hartree energy. We therefore make use of reduced range and padding for all further calculations, whilst restricting the resultant error in the total energies to be less than $10^{-3}$ eV.

\subsection{Polar molecules}

The test cases of a charged and dipolar system confirm that the cutoff interaction removes power-law convergence and indicate that it can yield some savings compared to finite-grid calculations. However, in the case of the NaCl bond, the energy difference being measured is rather small (around 0.2 eV), while the system itself consists of an extremely large dipole (calculated value of 8.2 D) in a fairly small box. As an example of a more realistic system of interest we calculated the structure of the molecule dimethylsulphoxide (DMSO), shown schematically in Fig. \ref{DMSOpic}, which still has a relatively large dipole (around 3.9 D), using the normal supercell approach and the cutoff interaction. Both calculations were converged to within 0.01 eV per atom, and to within 0.01 eV per atom of each other, for a cubic box size of 8 \AA\ and an energy cutoff of 650 eV. As shown in Table \ref{table:DMSO}, by the time the calculation is converged to this degree, the structures obtained already agree to within 0.002 \AA\ for bond lengths and $0.3^{\circ}$  for bond angles. This is a surprising result given the large dipole moment of DMSO, and the relatively small distance by which the molecules in the supercell calculation are separated.

There is also some interest in the calculation of molecular dipole moments using DFT.\cite{rashin} We found that the calculated dipole showed a greater proportional difference between methods than the structure, as a result of neither calculation giving a particularly well-converged dipole for box sizes of less than 12 \AA, where the dipoles agree within 0.1 D. This is because the dipole is far more sensitive to small movements of charge density from one side of the supercell to the other than the total energy.   

Use of the fully unaliased cutoff interaction made the calculation around five times slower than the unmoderated supercell calculation for DMSO in the 8 \AA\ box. The extra computation time is used almost entirely in the FFT of the padded density. Fully exploiting the reduced padding and range, by treating them as additional convergence parameters, as discussed earlier, resulted in a cutoff calculation now approximately twice as slow. 

The cutoff interaction method (or a finite-grid calculation) does not rely on any particular choice of supercell geometry. In the case of an infinite cubic lattice of point dipoles, the dipole-dipole interaction automatically cancels. Makov and Payne derived an energy correction for the macroscopic field imposed by PBCs and thus obtained the energy of the infinite cubic array which, due to the dipole-dipole cancellation, converged to the result for the isolated system as order $L^{-5}$. For non-cubic boxes the dipole-dipole interaction does not cancel and larger errors in a supercell calculation might be expected. The use of an artificially cubic supercell for elongated molecules in a supercell calculation, if needed, would involve expensive additional vacuum.

We therefore next considered the case of an elongated molecule in a non-cubic supercell. We performed a geometry optimisation for the molecule {\it p}-nitroanisole, with structure as shown in Fig. {\ref{pnapic}}. The box size for our converged calculation was $14 \ \mathrm{\AA} \times 10 \ \mathrm{\AA} \times 6 \ \mathrm{\AA}$, and an energy cutoff of 650 eV was needed. Representative results obtained for bond lengths and angles are shown in Table {\ref{table:PNA}}. Once again the similarity is striking, with bond lengths agreeing within 0.001 \AA\ and bond angles within $0.1^{\circ}$. Thus the structure of {\it p}-nitroanisole is described equally well by the supercell technique despite the non-cubic geometry (and even though the molecule has a somewhat larger dipole, calculated value 5.2 D, than DMSO). Total energies with and without interaction between molecules also agreed to well within 0.01 eV per atom. We emphasize that (even for these organic molecules chosen specifically for their atypically large dipole moments),  in order to obtain this agreement between supercell and finite results, the supercell was only as large as was required for convergence of the energy of the isolated system i.e., large enough to accommodate the charge density of the molecule. 

\subsection{Charged clusters}

As outlined earlier, supercell calculations for charged systems require the energy to be corrected by the insertion of a neutralising background, the Madelung correction, and, if necessary, the Makov-Payne correction, depending on the degree of accuracy required. The first two terms affect only the $\bG=0$ component of the total potential. This means that an unmoderated supercell calculation can in fact be carried out for a structure determination (because for a fixed number of electrons the error in total energy due to the first two terms is geometry independent). Any remaining errors in the resulting structure would be caused by interaction effects of the order addressed by the Makov-Payne correction.

We therefore carried out a structure determination for the ${\mathrm{Na}_{2}}^{+}$ cluster. We obtained a converged bond length with the cutoff interaction of 3.513 \AA, and with the supercell method of 3.512 \AA, in a box with dimensions $12 \ \mathrm{\AA} \times 10 \ \mathrm{\AA} \times 10 \ \mathrm{\AA}$, the smallest box size at which either calculation was well converged. In this case also, interaction effects (beyond the straightforward interaction of a lattice of charges which can be easily removed) have very little effect on the structure determination. 

\subsection{Coulomb interaction in finite-grid calculations}

In a PBCs calculation the Coulomb interaction is necessarily defined in reciprocal space - although the interaction is infinitely ranged both in real and reciprocal space, components of the density only extend out to some cutoff in reciprocal space, allowing in principle an exact determination of the Hartree energy.

In a finite-grid calculation the Coulomb interaction, which is now only required over some limited real space volume, is usually defined in real space as $1/r$. However, when viewed as a cutoff interaction in a periodic system, it is possible to use forms specified analytically in either real or reciprocal space.

As an example, we performed calculations for a Mg atom, using a Coulomb interaction cut off spherically at radius $R= \surd 3 L$ in real space, with appropriate padding, to determine the Hartree energy. This cutoff was imposed first by taking the analytical Fourier transform to reciprocal space,

\begin{equation}
V(\bG)= \frac{1 - {\mathrm cos}(|\bG|R)}{|\bG|^2},
\end{equation}
and then in real space, taking $1/r$ within the sphere and zero outside. Clearly the FFT of each should give the other, but only in the limit of a large energy cutoff (i.e., fine Fourier transform grid).

Convergence of the resulting total energies is shown in Fig. \ref{finint}. It can be seen that in fact convergence is much improved by using a cutoff interaction specified analytically in reciprocal space, rather than using $1/r$ with averaging of the singularity at the origin.

\section{Discussion}

Our calculations using the cutoff Coulomb interaction allow for detailed and unambiguous assessment of the effects caused by interaction between supercells. As the errors in energy caused by such interaction effects are relatively small, it is a crucial aspect of our work that we are able to retain the same pseudopotentials, exchange-correlation approximation, and basis set for the supercell and cutoff calculations. Although image interaction has often been cited as an argument against the supercell method, these factors have generally prevented accurate or conclusive comparisons, and the ability to perform such a study is the main motivation for our approach.

We firstly make two points regarding the implications of our results with the cutoff Hamiltonian for finite-grid calculations.  The savings we obtain through the reduced interaction range and padding verify that, although it is important to converge the exponential tail of the density for the total energy, the contribution to the Hartree energy from the interaction of these tails across one cell or between adjacent cells is not as significant. This also explains why the calculations of Onida {\it et al.} \cite{gwclusters} obtained well-converged results without employing padding, and without needing to fully separate clusters in different unit cells. This procedure is certainly worth considering in finite-grid calculations, although the proportional saving will be less for large systems (where the density decays exponentially only over a smaller proportion of the cell).

Implementation of the Coulomb cutoff in real and reciprocal space showed a marked difference in convergence with respect to energy cutoff. This is because the reciprocal space form treats the singularity in $1/r$ analytically. The slow convergence seen in the total energy with the real space interaction is simply a result of the need to sample $1/r$ very finely, and does not relate to the energy cutoff required to converge the density, which is far lower. As a result a cutoff interaction defined in reciprocal space (and then transformed to real space if necessary) should offer improved convergence for any finite computation of the Hartree energy.

The results of our calculations, especially for the dipolar molecules, reveal a very high level of accuracy in the structures and energies yielded by the supercell method. In particular, the common assertion that additional vacuum is required proves not to be the case for the charged cluster ${\mathrm{Na}_{2}}^{+}$ or for the molecules DMSO and {\it p}-nitroanisole, even though both have large dipoles. The excellent agreement seen in energies and structures suggests that the additional effort involved in a fully finite calculation is unlikely to be necessary for a wide range of clusters and molecules, as these errors are considerably smaller than the differences given by choice of pseudopotential or exchange-correlation approximation. This conclusion is consistent with the supercell calculations on $\mathrm{Cu}_{n}^{-}$ clusters by Massobrio {\it et al.}\cite{carclusters} who, using a positive jellium background, found that the size of the remaining error caused by interaction, as evaluated by the Makov-Payne formula, was already negligible compared to total-energy differences.

We therefore suggest that the most efficient method for determining structures and energies of clusters and molecules is to perform a conventional band structure calculation, converged with respect to the supercell dimensions. For charged systems or dipoles on a cubic lattice, involving, for example, very small energy differences, or very large dipoles, corrections to the energy can be assessed using the Makov-Payne terms. We note, though, that this approach for dipolar molecules can only be used in conjunction with cubic supercells. Another option is to restart from the optimized geometry with the cutoff interaction. We re-iterate though that, as we have found errors related to interaction to be less than 0.01 eV per atom in realistic cases, we expect that an uncorrected supercell calculation should nearly always suffice.

Most finite calculations are undertaken with Gaussian basis sets. In the absence of any adverse effects from enforced periodicity, plane-wave basis sets offer some advantages for {\it ab initio} calculations of clusters and molecules. Results of calculations undertaken with Gaussian basis sets are often dependent on the choice of basis set, as a result of being difficult to converge with respect to basis set size: there are also problems associated with lack of basis set orthonormality and description of diffuse occupied states. Plane waves offer a simple and systematic convergence procedure, and also facilitate straightforward computation of forces for molecular dynamics, and there has therefore been considerable interest in the applicability of plane-wave basis sets for first-principles chemical calculations for molecules\cite{rappe} and ions.\cite{ireta} Our work provides clear evidence that interaction between supercells provides no obstacles to a plane-wave approach.

Interaction effects are also potentially an issue in calculations of infinite systems such as polymers or surfaces, where a finite-grid calculation is not possible. It often proves necessary to include several layers of vacuum in supercell slab calculations of surfaces,\cite{7x7payne,7x7jo} for example. In the light of our results it seems possible that this may be purely the amount of vacuum needed to describe the exponential tail of the density, rather than an electrostatic effect. This would mean that no larger a supercell would be needed for the supercell method than would be required by using localized orbitals, although the differing geometry and larger polarizability do allow for potentially larger (and physically different) errors in these cases.

Similarly, in order to simulate defects in solids a very large supercell may be needed. Makov and Payne developed an energy correction for this case, based on earlier work by Leslie and Gillan,\cite{gillan} but our findings also raise the question of whether differences from the answer for the single defect in an infinite crystal are dominated by the need to describe the physical environment of the defect over medium range length scales or indirect electrostatic effects, rather than by direct electrostatic interaction between defects. Poor convergence because of chemical interactions is consistent with the finding that convergence with respect to supercell size depends on {\bf k}-point sampling.\cite{makov2}

The Makov-Payne correction to the total energy for dipolar systems in the case of cubic supercells accounts for much of the (small) difference between the PBCs calculation and the cutoff interaction, as was shown in Fig. \ref{NaCl}. The correction has the effect of reinstating the macroscopic field of the infinite array of supercells (which has been removed by the depolarizing field implicit in the use of PBCs), and calculates the corresponding energy. The remaining difference between the corrected energy and the isolated system is the interaction energy of the lattice. As the energy of a cubic lattice of point dipoles is zero, the corrected energy per cell should then give a good approximation to the energy of the finite system. We verified explicitly that the remaining (very small) discrepancy between the Makov-Payne-corrected energy and the converged result for the finite system could be accounted for almost entirely by the energy of a classical lattice of appropriate finite-length dipoles. It is also useful to observe that, although strictly the Makov-Payne correction requires a modification of the electrostatic energy functional and accordingly the effective potential, simply applying the correction using the density from a conventional supercell calculation actually provides a good estimate of the total-energy error.

For an infinite non-cubic lattice of dipoles, we cannot generally expect the interaction energy to be zero. However, for the {\it p}-nitroanisole molecule we studied, the interaction energy of the lattice (again evaluated for finite length dipoles) can be seen to have the opposite sign of the correction for the depolarising field, thus explaining why the supercell calculation for this system also differed only slightly from the cutoff calculation.

\section{Conclusions}

We have used a modified Coulomb interaction, cut off in real space, to allow calculations for finite systems to be undertaken within periodic boundary conditions, with no interaction between supercells.  We have demonstrated that, in any cases where a calculation for an isolated system should prove necessary, the cutoff Coulomb interaction technique can provide considerable efficiency gains compared with a finite-grid calculation by allowing a degree of aliasing in the Hartree integral without affecting the overall answer. We have also found that convergence with respect to energy cutoff of finite-grid calculations can be improved by defining the Coulomb interaction analytically in reciprocal space.

Detailed comparison of calculations performed for charged clusters and molecules with large dipoles shows that the supercell method gives very accurate energies and structures compared with the isolated system, meaning that the extra expense of finite-grid calculations is unnecessary in the large majority of cases. This shows that plane-wave basis set calculations are applicable to a wide range of chemical and biochemical problems without the need for artificially large supercells or corrections for supercell interaction.

\section*{Acknowledgements} 

We acknowledge useful discussions with Dr. M. J. Rutter, Dr. M. M. Rieger and R. E. Brown, and the support of the EPSRC.

\bibliographystyle{prsty}

\begin{thebibliography}{10}
\bibitem{kenote}
We actually use the kinetic energy for the continuous periodic system
  (evaluated in reciprocal space), but this equals the kinetic energy of the
  finite system in the limit that the density falls exponentially to zero at
  the box edges, as is required for convergence.
\bibitem{singnote}
This averaging can be easily performed numerically using a finer, offset,
  sub-grid for the volume element containing the origin.
\end{thebibliography}

\scalegraphlabel{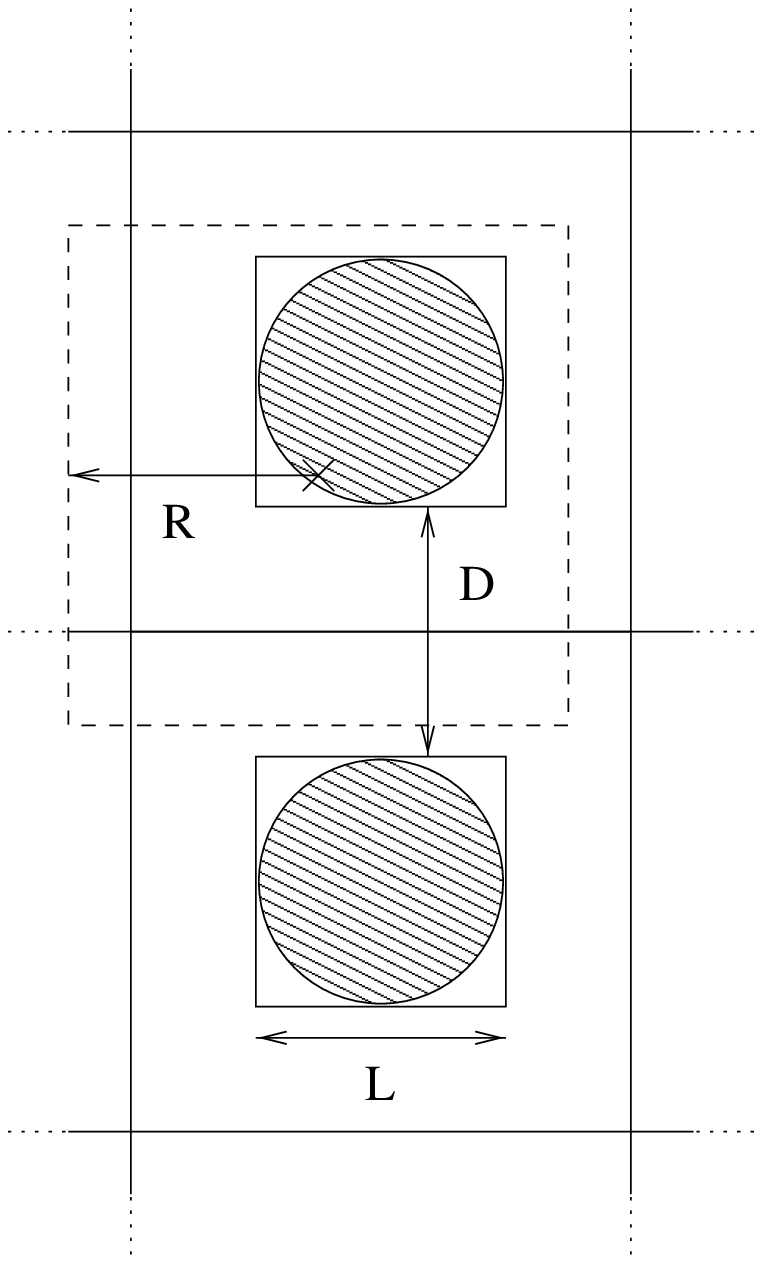}{5cm}{Schematic illustration of padding procedure for calculating the Hartree energy. Electronic density in a unit cell of side $L$ \ is padded by a distance $D$ \ and interacts over a box with range parameter $R$ \ (see text).}{padding}

\scalegraphlabel{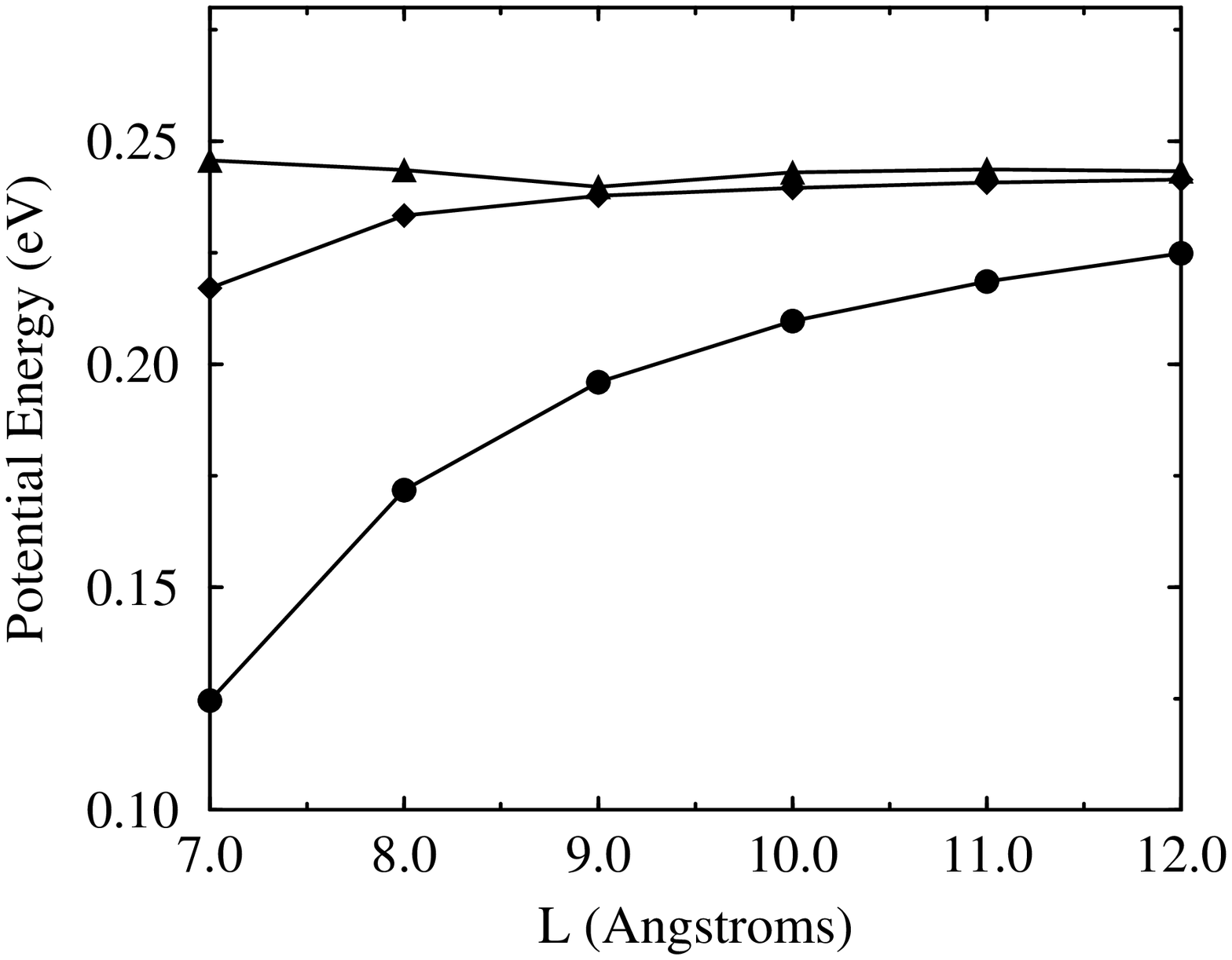}{7cm}{Energy of a NaCl bond with an extended bond length relative to equilibrium as a function of box size $L$, from supercell method (circles), corrected by Makov-Payne functional (diamonds), and using the cutoff Coulomb interaction (triangles).}{NaCl}

\scalegraphlabel{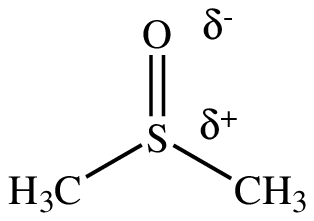}{3cm}{Structure of DMSO, which has a dipole as shown.}{DMSOpic}

\scalegraphlabel{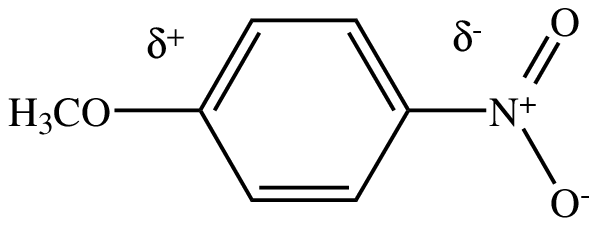}{5.7cm}{Structure of {\it p}-nitroanisole, with dipole as indicated.}{pnapic}

\scalegraphlabel{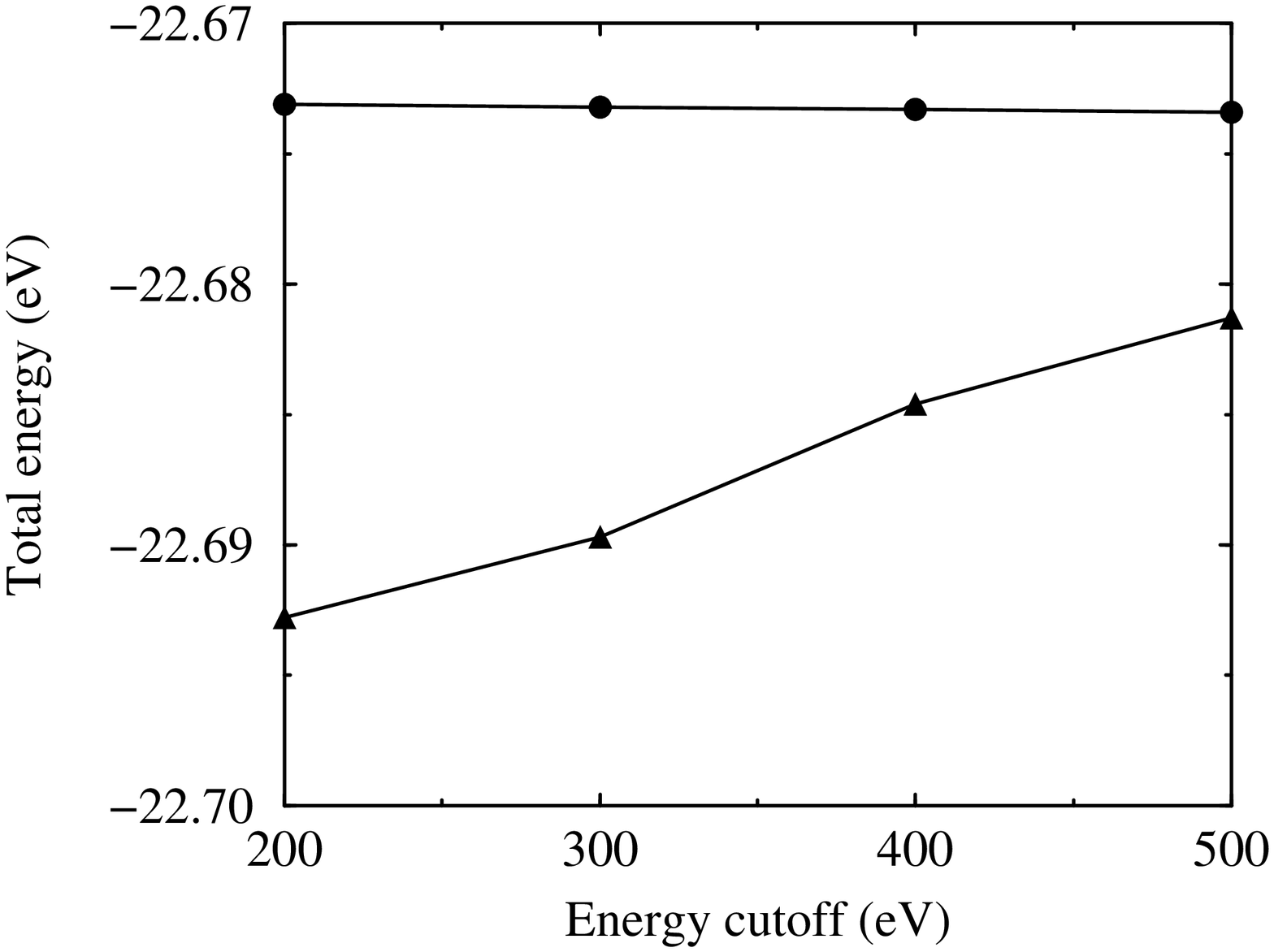}{7cm}{Total energy of Mg atom, calculated with cutoff Coulomb interaction, defined in real space (triangles) and reciprocal space (circles). }{finint}

\begin{table}
\caption{Bond lengths and bond angles for the molecule DMSO in a cubic box of side 8 \AA\ calculated using the usual supercell technique and with cutoff electrostatic interaction.}
\tabcolsep=0.3cm
\noindent
\begin{tabular}{r|cccccc}
& S-O & S-C & C-S-C & C-S-O \\
\hline
\\
Supercell & 1.433 \AA & 1.785 \AA & $95.8^{\circ}$ & $106.7^{\circ}$   \\
Cutoff & 1.431 \AA & 1.786 \AA & $96.1^{\circ}$ & $106.6^{\circ}$  \\
\end{tabular}
\label{table:DMSO}
\end{table}

\begin{table}
\caption{Bond lengths and bond angles for the molecule {\it p}-nitroanisole in a $14 \ \mathrm{\AA} \times 10 \ \mathrm{\AA} \times 6 \ \mathrm{\AA}$ box calculated using the usual supercell technique and with cutoff electrostatic interaction.}
\tabcolsep=0.3cm
\noindent
\begin{tabular}{r|cccccc}
& N-C & O-C & N-O & O-N-O \\
\hline
\\
Supercell & 1.485 \AA & 1.267 \AA & 1.078 \AA & $125.7^{\circ}$    \\
Cutoff & 1.485 \AA & 1.266 \AA & 1.078 \AA & $125.7^{\circ}$  \\
\end{tabular}
\label{table:PNA}
\end{table}

\end{document}